\documentstyle[12pt,epsf]{article}
\textheight 9in  \topmargin -.4in   \def\baselinestretch{1.4}
\def\Eq{\begin{equation}}	\def\End{\end{equation}}
\def\Eqa{\begin{eqnarray}}	\def\Enda{\end{eqnarray}}
\def\Endl#1{\label{#1} \End}	\def\Endla#1{\label{#1} \Enda}
	
		\def\to{\!\rightarrow\!}
\def\etal{{\it et.al.}}	\def\ie{{\it i.e.}}

\def\frac#1#2{{\textstyle {#1 \over #2}}}

\def\dspace#1{\renewcommand{\baselinestretch}{#1} \large\normalsize}
\def\dspace#1{\relax} % Uncomment this for single-spacing
\def\Sum{ \sum_{X \ne H}\!\!\!\!\raisebox{.8ex}{ {}'} }
\def\lesim{ \raisebox{-0.5ex}{$\stackrel{\textstyle <}{\sim}$} }
\def\gesim{ \raisebox{-0.5ex}{$\stackrel{\textstyle >}{\sim}$} }
\def\eps{\epsilon}
\def\btorhop{$\overline B^0 \to \rho^+ l^- \overline \nu$}
\def\btorho{$\overline B \to \rho l \overline \nu$}
\def\btopip{$\overline B^0 \to \pi^+ l^- \overline \nu$}
\def\btopi{$\overline B \to \pi l \overline \nu$}
\def\btods{$\overline B \to D^* l \overline \nu$}
\def\btod{$\overline B \to D l \overline \nu$}
\def\Lam{\bar \Lambda}
\def\lo{ \lambda_1}
\def\lt{ \lambda_2}

\def\OMIT#1{}
\def\ie{{\it i.e.}}
\def\etal{{\it et al.\/}}

\def\CO{{\cal O}}
\def\frac#1#2{{#1\over#2}}

\def\np#1#2#3{\NP{\bf B#1} (#2) #3}
\def\pl#1#2#3{\PL B {\bf #1} (#2) #3}

\def\pr#1#2#3{\PR{\bf #1} (#2) #3}

\begin{document}
\begin{titlepage}
\rightline{UCSD-TH-96-20}
\rightline{hep-ph/9607418}
\rightline{July 1996}
\vskip.5in
\begin{center}

{\Large \bf
Bounding Heavy Meson Form Factors Using Inclusive Sum Rules
}
\vskip.2in
{
  {\bf C. Glenn Boyd}\footnote{\tt gboyd@ucsd.edu}\\
\vskip.2cm
  {\bf  I.Z. Rothstein}\footnote{\tt ira@yukawa.ucsd.edu}\\
  \vskip.2cm {\it Department of Physics}\\
  \vskip.2cm {\it University Of California, San Diego}\\
  \vskip.2cm {\it La Jolla, California 92093-0319}}
\vskip.2cm
%PACS-96 numbers: 
\end{center}

\vskip.5in
\begin{abstract}
\dspace{1.4}

We utilize inclusive sum rules to construct both upper and lower bounds on the 
form factors
for $\overline B \to D, D^*, \rho, \pi,\omega,  K$ and $ K^*$ semi-leptonic
and radiative decays. We include the leading nonperturbative $1/E$ corrections 
and point out cases when $\alpha_s$ corrections are equally important.
We compute the $\alpha_s$ correction to the lower bound on the 
$\overline B \to D^*$ form factor $f(w)$ at zero recoil, thereby constraining
its normalization $f(1)$ to within $6-8\%$ of the upper bound.
We show that the $\overline B \to \rho$ form factor $a_+$ is suppressed
at small momentum transfer by either a factor of $1/E$ or $\alpha_s$. 
These bounds can be used to rule out phenomenological
models as well as to determine values for the CKM matrix elements once radiative
corrections are included. 
 
\par
\end{abstract}
\end{titlepage}
\dspace{2.0}
\setcounter{footnote}{0}

% ======================= TEXT BEGINS ==========================
\section{Introduction }

While there has been much progress in calculating inclusive decay rates
~[1-6]\ of
heavy mesons, exclusive rates have still not been tamed within the confines
of a first principles calculation. Consequently, the phenomenology of exclusive
decays has been relegated to the realm of models which, while quite useful on 
the
qualitative level, leave much to be desired when it comes to quantitative 
issues.
For instance, the CKM matrix element $V_{ub}$ is still only known to within a  
factor of two, because present extractions are based on model considerations. 
Inclusive 
techniques are plagued by large corrections in the theoretical calculations
\cite{ar}, and thus it seems that we  have no recourse but 
to try to tame the exclusive rate. Given that at present, we cannot calculate 
the form factors themselves, we do the next best thing, which is to {\it bound} 
them.

In reference \cite{BSUV2}, the equivalence of hadronic and partonic expressions 
for inclusive decay rates was 
used to derive sum rules analogous to those developed 
for deep inelastic scattering. These sum rules apply to heavy-heavy as well as 
heavy-light quark transitions, as long as the energy of the final state hadron 
is large compared to the QCD scale. An  explicit upper bound on the \btods\ 
matrix
element at zero recoil was presented (although radiative corrections 
significantly
weaken this bound~\cite{GKLW}) in \cite{BSUV2}. In this paper, we use these 
inclusive sum
rules to compute explicit bounds on individual heavy-heavy form factors at 
arbitrary momentum transfer, and heavy-light form factors at sufficiently small 
momentum transfer. In particular we bound form factors for the
transitions  $\overline B \to H l \overline \nu$, where $H$ can be a
$D, D^*, \rho, \omega,$ or $\pi$ meson, and $\overline B \to H l \overline l$
(or $\overline B \to H \gamma$), where $H$ can be a $K^*$
or $K$ meson. We show how to compute not only upper bounds, but {\it lower}
bounds as well, and present the explicit bounding functions.
Phenomenological issues like the extraction of $V_{ub}$ will be addressed in 
a subsequent publication, since such analyses require the inclusion of possibly 
large 
radiative corrections that are not included in the present article.

\section{Constructing Sum Rules}

Consider the semi-leptonic decay of a $B$ meson to a hadron 
$H$ through a vector $V^\mu = \overline q \gamma^\mu b$ or
axial $A^\mu = \overline q \gamma^\mu \gamma_5 b$ flavor-changing
current. Quark-hadron duality permits us to reliably calculate the inclusive
rate, after the requisite smearing over invariant mass \cite{PQW}, 
in terms of partonic
kinematic variables. The exclusive rates on the other hand, are not calculable
from first principles  and must be parameterized in terms of form factors.
Equating the calculable inclusive rate to the sum over exclusive modes leads
to the sum rules which will be utilized in this paper.
The sum rules are derived by noting that the time-ordered product
of two currents between $B$ mesons with four-velocity $v$,
\Eqa
T^{\mu\nu}(v\cdot q, q^2) &=& -i\int d^4 x\ e^{-i q \cdot x} \langle B(v)| 
T\left(
J^{\mu\,\dagger}\left(x\right) J^\nu\left(0\right)\right)
\left|{B(v)}\right\rangle\nonumber\\ &\equiv&-g^{\mu\nu} T_1 +
v^{\mu}v^{\nu} T_2 - i \epsilon^{\mu\nu\alpha\beta} v_\alpha q_\beta
T_3 + q^\mu q^\nu T_4\nonumber\\ && + \left(q^\mu v^\nu + q^\nu
v^\mu\right) T_5,
\Endla{tdef}
can be expressed as either a sum over hadronic or partonic intermediate states.
The former expression contains the matrix elements $ \left\langle H \left| J
  \right| B \right\rangle $ of interest, while the latter may be
expanded~\cite{chay}\ as an operator product expansion (OPE)  in the
heavy quark effective theory\cite{hqet}. 
Both the hadronic and OPE-based expressions for the time-ordered
product $T^{\mu\nu}$ may be analytically continued to
complex $v\cdot q$, holding the three-momentum
$q_3 = |\overrightarrow q|$ fixed. In terms of the variable
\Eq
\eps = M_B - E_H - v\cdot q,
\Endl{epsdef}
where $E_H = \sqrt{ M_H^2 + q_3^2}$ is the $H$ meson energy 
and $M_B$ is the $B$ meson mass,
$\: T^{\mu\nu}$ has two branch cuts along the real epsilon axis: a 
``local'' cut for $\eps \ge 0$ and a ``distant'' cut for $\eps \le  -2 E_H$. 
Far from these cuts, the OPE-based expression $T_{\mu\nu}^{OPE}$ should 
reliably approximate the hadronic one.

Contracting with an arbitrary four-vector $a^\mu$ and equating the 
hadronic sum over states to the OPE-based calculation gives 
\Eqa
&&{ | \langle H| a\cdot J | B \rangle |^2 \over 4 M_B E_H \eps }
+ \Sum { | \langle X| a\cdot J | B \rangle |^2 \over 4 M_B E_X
(\eps + E_H -E_X)} \nonumber \\
&&- \sum_X (2\pi)^3 \delta^{(3)}(\overrightarrow p_X
          - \overrightarrow q) { | \langle B| a\cdot J | X \rangle |^2 \over 4 
M_B E_X
(\eps + E_H +E_X -2 M_B)} 
= a^{\mu*} T^{OPE}_{\mu\nu} a^\nu .
\Endla{aTa}
The first two terms represent the local cut, while the third term, which
sums over states $X$ containing one $q$ and two $b$ quarks, represents the
distant cut.
The sum over states contains the usual phase space integration
$\int d^3p/ (2 E)$ for each particle, while $\Sigma_{X \ne H}'$ is shorthand for
\Eq
\Sum \equiv \sum_{X \ne H}\, (2\pi)^3 \delta^{(3)}
(\overrightarrow p_X
          + \overrightarrow q).\nonumber
\Endl{sigdef}

Eq.~\ref{aTa}\ is derived by assuming $\eps$ is real, then analytically
continuing to complex $\eps$. Following the procedure outlined in references
\cite{BSUV2} and \cite{GKLW}, we integrate in $\eps$ along a contour
that encloses only the local branch cut while remaining far from either
cut (except at $\eps \to \infty$, where local duality is expected to work
well). The $ | \langle B| a\cdot J | X \rangle |^2 $ term in 
Eq.~\ref{aTa} will then give a vanishing contribution. 
To ensure convergence, we multiply $a^\dagger T a$ by
a smooth weight function $W_\Delta(\eps)$ satisfying 
$W_\Delta(0)=1$, $W_\Delta(\eps) \to 0$ for $\eps << \Delta$, and
$W_\Delta(\eps) > 0$ for $\eps$ real. $\Delta$ acts as an ultra-violet
cutoff which  serves to damp the contribution from excited states.
The result of this 
integration, $\int W_\Delta(\eps)\ d\eps$, is the zeroth moment rule
\Eq
{ | \langle H| a\cdot J | B \rangle |^2 \over 4 M_B E_H }
+ \Sum { | \langle X| a\cdot J | B \rangle |^2 \over 4 M_B E_X }
 W_\Delta(E_X -E_H) = \int d\eps\ W_\Delta(\eps)\ a^{\mu*} T^{OPE}_{\mu\nu} 
a^\nu .
\Endl{zmoment}
The positivity of $| \langle X| a\cdot J | B \rangle |^2$ gives an immediate
upper bound on the magnitude of the combination of form factors entering
$\langle H| a\cdot J | B \rangle$.

Integrating $\int W_\Delta(\eps)\, \eps\ d\eps$ gives the sum rule for the first 
moment,
\Eq
\Sum {(E_X - E_H) | \langle X| v\cdot J | B \rangle |^2 \over 4 M_B 
E_X } 
W_\Delta(E_X -E_H) = \int\eps\ d\eps W_\Delta(\eps)\ a^{\mu*} T^{OPE}_{\mu\nu} 
a^\nu .
\Endl{firstmoment}
This leads to lower bounds on form factors by noting that, if
$E_1$ is the energy of the first resonance more massive than $H$,
\Eq
(E_1 - E_H) \Sum { | \langle X| a\cdot J | B \rangle |^2\over 4 M_B 
E_X }
          W_\Delta(E_X -E_H)  \le
 \ \ \Sum {(E_X - E_H) | \langle X| a\cdot J | B \rangle |^2 \over 
4 M_B E_X }
          W_\Delta(E_X -E_H)  .
\Endl{voltrick}
We neglect the contribution of multi-particle states with energies less 
than that of the first excited resonance. The contributions of such 
states are suppressed by both phase space and large-$N_c$ power counting and 
moreover, are empirically negligible (e.g., $D \to K^* \mu \nu$ versus
$D \to K \pi \mu \nu$).

Substituting Eq.~\ref{voltrick}\ into Eq.~\ref{firstmoment}\ 
provides an upper bound on the contribution of excited states 
to the zeroth moment rule Eq.~\ref{zmoment}. This in turn implies a lower bound
on the hadronic matrix element $\langle H| a\cdot J | B \rangle$.  We therefore
have both the upper and lower bounds
\Eq
\int d\eps\ W_\Delta(\eps)\ a^{\mu*} T^{OPE}_{\mu\nu} a^\nu \ge 
{| \langle H| a\cdot J | B \rangle |^2 \over 4 M_B E_H }\ge 
\int d\eps\ W_\Delta(\eps)\ a^{\mu*} T^{OPE}_{\mu\nu} a^\nu 
 \biggl[ 1 - { \eps\over E_1 - E_H} \biggr]  .
\Endl{generalbd}
Eq.~\ref{voltrick}\ was previously used for deriving the Voloshin bound on the 
slope of the \btods\ form factor at zero recoil\cite{Volo}. The bounds derived 
here apply to
the normalizations of form factors, rather than the slopes, and may be used
away from zero recoil as well.
We may now use Eq.~\ref{generalbd} to bound the form factor of our choosing 
by
appropriately selecting the four-vector $a_\mu$ and current $J_\mu$.
Furthermore, variation of  $q_3$ leads to constraints 
over the entire physical range of momentum transfer $q^2$.
When the first moment of $a^\dagger T a$ is small, the upper and lower
bounds are close to each other, and the form factor is tightly constrained. 
Naturally, this is the most interesting kinematic region to consider, but
care is required since higher order terms become important.

There are several expansion parameters implicit in Eq.~\ref{generalbd}. The
OPE result contains powers of $\Lambda /  m_b$ from matching to the heavy quark
effective theory, $ \Lambda / 2 E_q $ from expressing the time-ordered product 
as 
a sum of local operators and $\Lambda / \Delta$ from derivatives of the weight 
function
$W_\Delta$, where $\Lambda$ is a typical hadronic energy scale. To the order
at which we work, the $\Lambda / \Delta$ terms can be eliminated by taking 
$\Delta \sim E_q$ and choosing a weight function whose 
first and second derivatives vanish at zero. Thus, 
$ \Lambda / 2 E_q $ is the limiting parameter and the bounds are only valid for 
sufficiently large energies, at least $E_q \gesim 1\, {\rm GeV}$, corresponding 
to 
small $q^2$.
For $\overline B \to \rho, \pi, \omega$, the maximum energy of the final hadron
is about $2.7~{\rm GeV}$, so the bounds can be valid over a substantial 
kinematic 
range, roughly given by $ 0 \leq q^2 \lesim 18 {\rm GeV}^2$.
Since our integration contour necessarily approaches either the local or distant
branch cut to within $E_H$, this requirement also enforces the local 
duality condition that the contour remain far from any cuts. 

In addition there are perturbative corrections that we expect, 
for $\Delta \sim  E_q$, to be the same order as the $1/2 E_q$ corrections. 
Schematically, the corrections to the first moment enter in the form
\Eq
 \Lambda + {\lo + \lt\over 2 E_q} + {\alpha_s(\Delta) \over \pi}\Delta 
       + \cdots ,
\End
where functions of $q_3$ and particle masses multiply each of the terms
above. When the leading $\Lambda$ term vanishes, both the $1/2 E_q$
terms presented in this paper and the uncalculated $\alpha_s$ corrections 
are dominant. The $\alpha_s$ corrections need to be calculated before
our lower bounds can be reliably applied in this kinematic region.

\section{ The Hadronic Side }
To apply the generic bounds Eq.~\ref{generalbd}\ to a specific
form factor, we must choose an appropriate current $J$ and four-vector $a^\mu$.  
The matrix elements 
for semi-leptonic decay of a $B$ meson into a pseudoscalar meson $P$ or
a vector meson $V$ may be parameterized as 
\Eqa
&&\langle P(p^\prime)\mid V^\mu \mid \bar B(p)\rangle =
\left(p+p^\prime\right)^\mu f_+ + \left(p-p^\prime\right)^\mu f_-,\\
\noalign{\smallskip} &&\langle V(p^\prime)\mid V^\mu
\mid\bar B(p)\rangle = i g \epsilon^{\mu\nu\alpha\beta}
\epsilon^*_{\nu}p^\prime_\alpha p_\beta,\\ \noalign{\smallskip}
&&\langle V(p^\prime)\mid A^\mu \mid\bar B(p)\rangle
= f \epsilon^{*\mu} +\left[
\left(p+p^\prime\right)^\mu a_+ + \left(p-p^\prime\right)^\mu a_-
\right] p\cdot \epsilon^*.
\Endla{param}
The states in Eq.~(\ref{param}) have the usual relativistic normalization of 
$2E$.
Contributions to decay rates from $a_-$ and $f_-$ are suppressed 
by the lepton mass and are therefore of less interest.

The tensor coefficients $T_i$ of the time-ordered product $T^{\mu\nu}$
receive contributions\cite{BGM}\ from the above matrix elements.
Decays to pseudoscalar mesons contribute 
\begin{eqnarray}\label{TP}
&&  T_1 = 0, \qquad \qquad T_2 = 2 f_+^2\ M_B {1\over \Delta_P},
 \qquad \qquad T_3 = 0,\nonumber
\\ &&T_4 = \left(f_+-f_-\right)^2\ {1\over 2 M_B\Delta_P},
 \qquad \qquad T_5 =  f_+\left(f_--f_+\right) {1\over \Delta_P},
\end{eqnarray}
while decays to vectors contribute 
\begin{eqnarray}\label{TV}
&&T_1 = \left[ g^2\left(p\cdot q^2 - M_B^2 q^2 \right) +
f^2\right] {1\over 2 M_B \Delta_V},\nonumber \\ \noalign{\medskip} &&
T_2 = \left[-q^2 g^2 + {f^2\over M_{V}^2} + 4
a_+^2\left(-M_B^2 + { \left(M_B^2-p\cdot q\right)^2\over
M_{V}^2}\right) + 4 f\ a_+\left( -1 + {M_B^2\over M_{V}^2} - 
{p\cdot q \over M_{V}^2}\right)\right] M_B\,
{1\over 2 \Delta_V}, \nonumber\\ \noalign{\medskip} &&
T_3 = g\, f\ {1\over \Delta_V}, \nonumber\\
\noalign{\medskip} &&
T_4 = \left[-g^2 M_B^2 + {f^2\over
M_{V}^2} + \left(a_+-a_-\right)^2\left( -M_B^2 +{\left(M_B^2-p\cdot
q\right)^2\over M_{V}^2}\right) \right.
\nonumber\\ &&\qquad\qquad\qquad + 2 f
\left.
\left(a_+-a_-\right){M_B^2-p\cdot q\over M_{V}^2}\right]
{1\over 2 M_B\Delta_V} ,\nonumber\\ \noalign{\medskip} &&
T_5 =\left[p\cdot q\ g^2 - {f^2\over M_{V}^2} + 2 a_+^2
\left(M_B^2 - { \left(M_B^2-p\cdot q\right)^2\over M_{V}^2}\right) +
f\ a_+\left( 1-3{\left(M_B^2-p\cdot q\right)\over
M_{V}^2}\right)\right. + \nonumber\\ &&\qquad\qquad\qquad\left. f\
a_- \left({\left(M_B^2-p\cdot q\right)\over M_{V}^2}-1\right) + 2
a_+\ a_- \left(-M_B^2 + { \left(M_B^2-p\cdot q\right)^2\over
M_{V}^2}\right) \right] \,
{1\over 2 \Delta_V}.
\end{eqnarray}
$\Delta_H$ (for $H=P$ or $H=V$) is the $H$ meson inverse propagator defined by
$\Delta_H = \left(p-q\right)^2 - M_H^2 = \eps\, (\eps + 2 E_H)$.
The contributions from decays to scalar or axial vector mesons are exactly 
analogous
to that of pseudoscalar or vector mesons, respectively, after interchanging
vector and axial vector currents $V^\mu \leftrightarrow A^\mu$.

Eq.s~\ref{TP}\ and \ref{TV}\ allow us to express the hadronic side of
the sum rule, involving $\langle H| a\cdot J | B \rangle$, in terms
of the form factors in Eq.~\ref{param}.
Isolating an individual form factor is now reduced to
making the appropriate choice for $a_\mu$.
It is convenient to go to the $B$ rest frame with the $z$ axis in the direction 
of $\overrightarrow q$, $q = (v\cdot q, 0,0,q_3)$. 
In this frame we may isolate the form factor $f\;$ ($g$)  by making the choice
$a = (0,1,0,0)$ and $J=V \;$ ($A$), which 
selects the sum rule $T_1^{hadronic} = T_1^{OPE}$.  
Since decays to scalars do not contribute to $T_1$,
the first excited resonance has spin/parity $J^P= 1^+$. 
These resonances are  $b_1(1235)$, $K_1(1270)$ and  $D_1(2420)$ 
for the transitions  \btorhop\ , $\overline B \to K^* l \overline l$ and 
\btods\, respectively.

Similarly, we may isolate the form factor $f_+$ via the choice
$a = (q_3, 0,0, v\cdot q)$ and $J=V$, leading to the  combination 
$a^\dagger T a = q^2 T_1 + q_3^2 T_2$. 
This combination has the advantage that no $J^P=1^-$ states
contribute, so the first excited resonance is again a $J^P= 1^+$ state.
Had we instead chosen $a = (q_3,0,0,M_B - E_\pi)$ to isolate $f_+$, the 
first excited resonance would have been the $\rho$, resulting in less 
stringent bounds. 

Isolation of the form factor $a_+$ requires the  use  of the heavy quark 
relation
$a_- = - a_+ (1 + \CO(1/m_b) )$. Since $1/m_b$ corrections are smaller 
than $1/E$ corrections, we can use this relation to eliminate $a_-$
from $a^\dagger T a$. Choosing $a = ( E_H, 0,0,-q_3)$ and $J=A$ then
selects
$
a^\dagger T a =
-M_H^2 T_1 + E_H^2 T_2 +  (M_B E_H - \eps E_H- M_H^2 )^2 T_4
+ 2 E_H (M_B E_H -\eps E_H - M_H^2) T_5,
$
which isolates $a_+$. The first excited resonance in this case can 
be a scalar $J^P= 0^+$ or an axial vector  $J^P= 1^+$. 
These states correspond to
 $a_0(980)$, $K_1(1270)$ and  $D_1(2420)$, for the transitions
\btorhop, $\overline B \to K^* l \overline l$ and  \btods respectively.

We may also consider the phenomenologically interesting decay $B^- \to 
\omega l \overline\nu$
by noting that both intermediate $\rho^0$ and $\omega$
states contribute to Eq.~\ref{tdef} when the external state is a charged
$B^-$ meson, but only $\rho^+$ contributes when the external
state is neutral. By using isospin to relate $\overline B^0 \to \rho^+$ and
$B^- \to \rho^0$ form factors, we can substitute the upper and 
lower bounds on \btorhop\ form factors, Eq.~\ref{generalbd}, into the sum rules
involving $\rho^0$ and $\omega$ intermediate states. This results in upper and 
lower bounds on form factors for $B^- \to \omega l \overline\nu$.

\section{The OPE Side}

Having fixed $J$ and $a$ to determine the hadronic side of the sum rule, 
we need to compute the OPE expression for
$a^\dagger T a$.  The zeroth order OPE result is simply the naive parton 
model, while the leading nonperturbative corrections can be written in terms 
of the parameters [2-5]
\Eqa
\lo &=& \left\langle
B(v) \right| \bar b_v {\left(iD\right)^2} b_v\left| B(v)
\right\rangle, \nonumber \\
\lt &=& -Z_b \left\langle B(v) \right| \bar b_v
{g\sigma^{\mu\nu}G_{\mu\nu}\over6} b_v\left| B(v)
\right\rangle,
\Endla{lamdef}
where $Z_b$ is a renormalization factor equal to unity at a scale
$\mu=m_b$.  The bottom quark mass may be eliminated by using the relation
\Eq
m_b = M_B - \Lam  + (\lo + 3\lt)/( 2 M_B) + \ldots. 
\end{equation}
The matrix
element $\lt = 0.12\, {\rm GeV}^2$ is determined by the $B^*-B$ mass
splitting, while $\lo$ and $\Lam$ may be extracted from inclusive
decay distributions\cite{GKLW2,FLS2}. 
 
The zeroth, first, and second moments of $T_1, T_2,$ and $T_3$ have been 
calculated
in reference~\cite{BSUV2}. We  also need the moments of $T_4$ and $T_5$
for a $b \to q$ changing axial current (the result for a vector current may 
be obtained  by making the replacing the final quark mass $m_q \to - m_q$).
Due to the mismatch between the definition of $\eps$ in terms of hadronic 
variables and the computation of the OPE in terms of partonic variables,
the OPE is an expansion about $\delta \equiv E_q - E_H + M_B - m_b$,
\Eq
     T_i = \sum_n { A^{(n)}_i \over (\eps - \delta)^{n+1} }.
\Endl{Adef}
The $A_i^{(n)}$ would be the $n^{th}$ moments of $T_i$ if we
defined as $\eps = m_b - E_q - v\cdot q$.  For $T_4$, they are given by
\Eq
A_4^{(0)}  =  {\lo + 3 \lt \over 3 M_B E_q^3},\qquad \qquad 
A_4^{(1)} = - {\lo + 3 \lt \over 3 M_B E_q^2},\qquad \qquad 
A_4^{(2)} = 0, 
\Endl{t4mom}
while for $T_5$, we have
\Eqa
A_5^{(0)} &=&  {-1/2 E_q} -{5\lt \over 4 E_q^3} - ({1\over 2 E_q^3}
       + {m_q^2\over 4 E_q^5})\lo ,  \nonumber \\
A_5^{(1)} &=&   ( {5\over 4 E_q^2} -  {5 \over 4 M_B E_q})\lt
 +( {1\over 2 E_q^2} + {m_q^2 \over 4 E_q^4} -{5\over 12 M_B E_q})\lo ,
                            \nonumber \\
A_5^{(2)} &=&  ({1\over 6 E_q} - {m_q^2 \over 6 E_q^3}) \lo .
\Endla{t5mom}
Higher moments will not contribute at this order. It is a simple matter to 
construct 
the moments of $T_i$ from $A_i$. The first moment, for example, is
$ \int \eps\, d\eps T = \delta A^{(0)} + A^{(1)}$. For \btod, $\delta$ 
may be set to zero when multiplying
higher order corrections $\lo, \lt$
as
$\delta \approx \Lam (w-1)/w$, where $w$ is the velocity transfer 
$v \cdot v'$. For \btorho, we keep factors
of $E_q - E_H$ multiplying such terms even though $M_H \sim \Lambda$
formally implies $E_q - E_H \sim \CO(\Lambda^2/E_q) $, because they are 
numerically important $M_H^2 >> \lo, \lt$. 

Given a four-vector $a^\mu$, we can now construct OPE expressions for the
moments of the sum rule combinations $a^\dagger T a$.
Using Eq.~\ref{generalbd} then leads to the bounds 
\Eqa
&&{E_q + m_q \over 2 E_q} 
 + ({1\over M_B^2} 
+ {2 m_q \over 3 M_B E_q^2 } + { m_q^2 \over E_q^4}) {m_q \lo \over 4 E_q}
          + ({3\over M_B^2} +{2 m_q\over M_B E_q^2} - {1\over E_q^2}){m_q \lt 
\over 4 E_q}
\nonumber \\ && \ge  {f^2 \over 4 M_B E_H} \nonumber \\
&& \ge  {1\over 2(E_1-E_H)} \biggl[ {E_q + m_q \over E_q} (E_1 
-E_q -\Lam)
\nonumber \\ && + ( {1 \over 3 E_q} - {1 \over 3 M_B} - {m_q\over 2 M_B^2} 
   +{m_q E_1 \over 2 M_B^2 E_q} + {m_q^2 \over 6 E_q^3} + {m_q^2 E_1 \over 3 M_B 
E_q^3}
  + {m_q^3 E_1 \over 2 E_q^5} ) \lo \nonumber \\ &&
   + ( -{1 \over 2 E_q} - {1 \over  M_B} - {3 m_q \over 2M_B^2}
   - {m_q E_1 \over 2 E_q^3} + {3 m_q E_1 \over 2 M_B^2 E_q} 
    + {m_q^2 E_1 \over M_B E_q^3} ) \lt \biggr]
\Endla{fbd}
for the final states $H = D^*, K^*,$ or $\rho^+$.
The upper and lower bounds on $ f^2 / (4 M_B E_H)$ in the above equation 
serve also to bound the vector form factor $g^2 M_B q_3^2 /(4 E_H)$ after
replacing $m_q \to - m_q$.  

The bounds on $a_+$, which involve higher moments, are given by
\Eqa
&& {E_H^2 + q_3^2 \over 2} -{E_H q_3^2 \over E_q} 
- {m_q M_H^2\over 2 E_q}  \nonumber \\
&& +\biggl[ {q_3^4 \over 3 M_B E_q^3}
    -{E_H q_3^2 \over 3 E_q^3} - {E_H^2 \over 3 E_q M_B} 
   -{M_H^2 m_q\over 4 M_B^2 E_q} - ({M_H^2 \over 6 M_B E_q^3}
 +{E_H q_3^2\over 2 E_q^5})m_q^2 - {M_H^2 m_q^3 \over 4 E_q^5}  \biggr]\lo
\nonumber \\
&& +\biggl[ {q_3^4 \over M_B E_q^3} -{E_H q_3^2 \over 2 E_q^3}
       - {E_H^2 \over M_B E_q} -({3 E_H^2 + q_3^2\over 4 E_q^3}
     +{3 M_H^2 \over 4 M_B^2 E_q})m_q - {M_H^2 m_q^2 \over 2 M_B E_q^3}
      \biggl] \lt \nonumber \\  
&\ge& \quad  {q_3^2 M_B M_H^2 a_+^2 \over E_H} \nonumber  \\ \quad \quad 
&\ge& \quad 
   {1\over E_1 -E_H} \Biggl\{ \biggl[ {E_H^2 + q_3^2 \over 2} -{E_H q_3^2 \over 
E_q}
- {m_q M_H^2\over 2 E_q}\biggr] (E_1 -E_q -\Lam) \nonumber \\
&+& \biggl[ {E_H^2+q_3^2\over 6 E_q} -{E_H q_3^2 + E_1 E_H^2\over 3 E_q M_B}
    +{2 E_H^2 + M_H^2 \over 6 M_B} - {E_H q_3^2 E_1 \over 3 E_q^3}
+{q_3^4 E_1 \over 3 M_B E_q^3} \nonumber  \\ \quad \quad
&+& {M_H^2 m_q \over 4 M_B^2}(1 - {E_1\over E_q}) 
 - ({E_1 M_H^2 \over 6 M_B E_q^3} - { E_H^2 + q_3^2 \over 12 E_q^3}
 +{ E_H E_1 q_3^2 \over 2 E_q^5}) m_q^2 - {E_1 M_H^2 m_q^3 \over 4 E_q^5}
\biggr]\lo \nonumber  \\ \quad
&+& \biggl[ {2 E_H^2 + M_H^2 \over 2 M_B}  +{ 2 E_H^2 + M_H^2 \over 4 E_q}
 -{E_1 E_H^2 + E_H q_3^2 \over E_q M_B} -{ E_H E_1 q_3^2 \over 2 E_q^3}
 + { E_1 q_3^4 \over M_B E_q^3}  \nonumber  \\ \quad
&-& ( {E_1 (3 E_H^2 + q_3^2) \over 4 E_q^3} -{3 M_H^2 \over 4 M_B^2}
     +{ 3 E_1 M_H^2 \over 4 M_B^2 E_q} )m_q - {E_1 M_H^2 m_q^2 \over 2 M_B 
E_q^3}
 \biggr]\lt  \Biggr\} .
\Endla{apbd}
For the sake of bookkeeping we  have retained the  $1/M_B$ terms even though 
the relation $a_- =- a_+$  
used on the hadronic side of this sum rule is valid only to $\CO(M_B^0)$.
When $m_q=0$, the zeroth order term in the upper bound, $(E_\rho -q_3)^2 $,
should naively be $\CO(q_3^2)$, but is actually 
$\CO( \Lambda^4/q_3^2)$ for $q_3 >> m_\rho \sim \Lambda$, so 
the $\lo,\lt$ terms are formally leading. Thus, $a_+$ is suppressed,
relative to naive expectations, by either $1/2 E_q$ or $\alpha_s$.

Bounds on $f_+$ for \btopip\ also involve higher moments since the 
four-vector $a = (q_3,0,0,v\cdot q)$ depends on $\eps$.  We find  
\Eqa
  &&   {E_q -m_q \over 2 E_q}(M_B -\Lam + m_q)^2 
 + \biggl[{M_B\over E_q} - \frac56 + ({1\over 2 M_B} -{1\over 12 E_q})m_q
  +{M_B \over 6 E_q^3}m_q^2 \nonumber  \\
&+& ({-M_B^2\over4 E_q^5} + {1\over 3 E_q^3}
     -{1\over 4 M_B^2 E_q} )m_q^3 + ({-M_B\over 2 E_q^5 } +{1\over 6 M_B 
E_q^3})m_q^4
      - {1\over 4 E_q^5}m_q^5 \biggr]\lo \nonumber  \\
 &+& \biggl[-\frac12 +{M_B\over E_q} + ({3\over 2 M_B} +{M_B^2\over 4 E_q^3}
      -{1\over 4 E_q})m_q -{M_B\over E_q^3}m_q^2
     -3 ({1\over 4 E_q^3} + {1\over 4 M_B^2 E_q})m_q^3 + {m_q^4\over 2 M_B 
E_q^3}\biggr]\lt
\nonumber  \\ \qquad &\ge&  { f_+^2 M_B q_3^2\over E_\pi}
\nonumber  \\ \qquad \qquad &\ge&
   {1\over (E_1 -E_H)}\Biggl\{ {(E_q -m_q) \over 2 E_q}(M_B-\Lam + m_q)^2(E_1 
-E_q -\Lam)
\nonumber  \\
&+&\biggl[5 E_q -5 E_1 -7 M_B +{6M_B E_1 + M_B^2\over E_q}
  +(-\frac32 +{3 E_1 -3 E_q\over M_B} +{4 M_B -E_1\over 2 E_q}) m_q \nonumber  
\\
 &+&({2 M_B E_1 +M_B^2\over 2 E_q^3} -{1\over M_B} +{1\over E_q})m_q^2
+({M_B +2 E_1\over E_q^3} +{ 3E_q -3E_1 \over 2 M_B^2 E_q}
      -{3 M_B^2 E_1 \over 2 E_q^5})m_q^3  \nonumber  \\
&+& ({M_B +2 E_1\over 2 M_B E_q^3} - {3 M_B E_1 \over E_q^5}) m_q^4 
     - {3 E_1\over 2 E_q^5}m_q^5 \biggr]{\lo\over 6} \nonumber  \\  
&+& \biggl[ 2 (M_B + E_q -E_1) +{4 M_B E_1 - M_B^2 \over E_q}
+( 5 + 6{ E_1 -  E_q \over M_B} -{2 M_B + E_1 \over E_q} +{M_B^2 E_1 \over 
E_q^3})m_q
\nonumber  \\ &-&({2\over M_B} + {1\over E_q} +{4 M_B E_1 \over E_q^3}) m_q^2
+ 3 ({1\over M_B^2} - {E_1\over E_q^3} -{E_1 \over M_B^2 E_q}) m_q^3
    +{2 E_1\over M_B E_q^3}m_q^4 \biggr] {\lt\over4} \Biggl\}. 
\Endla{fpbd}

For $B^- \to \omega l \overline\nu$, we present bounds
only for the form factor $f^{(B \to \omega)}$. Since these are derived
by combining two sum rules, they depend on the energy $E_1$ of the
first neutral $J^P=1^-$ resonance above the $\omega$, the $\Phi(1020)$,
as well as the energy $E_{b1}$ of the first charged $J^P=1^+$ resonance above 
the $\rho$, the $b_1(1235)$. Setting $m_q=0$ and $E_q=q_3$ 
gives the bounds 
\Eqa
&&{1\over 4 (E_{b1} - E_\rho)}\Biggl[ E_{b1} + q_3 - 2 E_\rho + \Lam 
 + \biggl( {1\over 3 M_B} - {1\over 3 q_3} \biggr)\lo
 +  \biggl( {1\over M_B} + {1\over 2 q_3} \biggr)\lt \Biggr]
\nonumber \\ && \ge
 {f^2_{ (B \to \omega)} \over 4 M_B E_\omega} 
\nonumber \\ && \ge
 {1\over 4 (E_1 - E_\omega)}\Biggl[ E_1 + E_\rho - 2 q_3 - 2 \Lam
   +  \biggl( {-2\over 3 M_B} + {2\over 3 q_3} \biggr)\lo 
 -  \biggl( {2\over M_B} + {1\over q_3} \biggr)\lt \Biggr] 
\Endla{bomega}
The upper bound on $f^{(B \to \omega)}$ is better than the
naive one in  Eq.~\ref{generalbd}\ for larger values of $q^2$, roughly
when $q_3 < (M_{b1}^2 - \Lam^2)/ (2 \Lam)$.  The lower bound is
only useful for large momentum transfer and rather small values
of $\Lam$.
    
\section{ Discussion }    

Let us consider the reliability of the bounds derived above.
We notice that the upper bounds rely only upon the zeroth moments 
and generally receive small corrections from $1/E$ nonperturbative terms. 
Perturbative $\alpha_s$ corrections should be similarly small, so most
of the upper bounds are trustworthy.
An exception is the upper bound on $a_+$. In this case
the zeroth moment is dominated by both the $1/2 E_q$ and $\alpha_s$ terms, 
and our result must be supplemented by a perturbative calculation. 
Even without such a calculation, 
we see that $a_+$ is dynamically suppressed for large $E_\rho$ , \ie\
for small momentum transfer $q^2 \lesim 18\, {\rm GeV}^2$.  
There have been attempts to calculate form factors such as $a_+$ within the 
confines of perturbative QCD utilizing Sudakov resummations to avoid 
the use of arbitrary cutoffs in the end point region, see \cite{akhster}.
These methods also predict an $\alpha_s$ suppression for small $q^2$,
but their normalization depends on unknown hadronic wavefunctions.

The lower bounds exhibit cancelations over much of the $q^2$ range and   
are most interesting when the first moment is small, of order
the $1/2 E_q$ terms, thus making the inclusion of short distance
corrections imperative.
Nevertheless, the $\CO(\alpha_s^0)$ formulas presented
here are a necessary first step and give a rough idea of how constraining
the lower bounds might be. Since the precise numerics are irrelevant without
the $\alpha_s$ corrections, we will only discuss the qualitative behavior
of some representative bounds.

%%%%%%%%%%%%%%%%%%%%%%%%%%%%%%%%%
\begin{figure}
\centerline{
\hfill 
\epsfxsize=0.4\textwidth
\epsffile{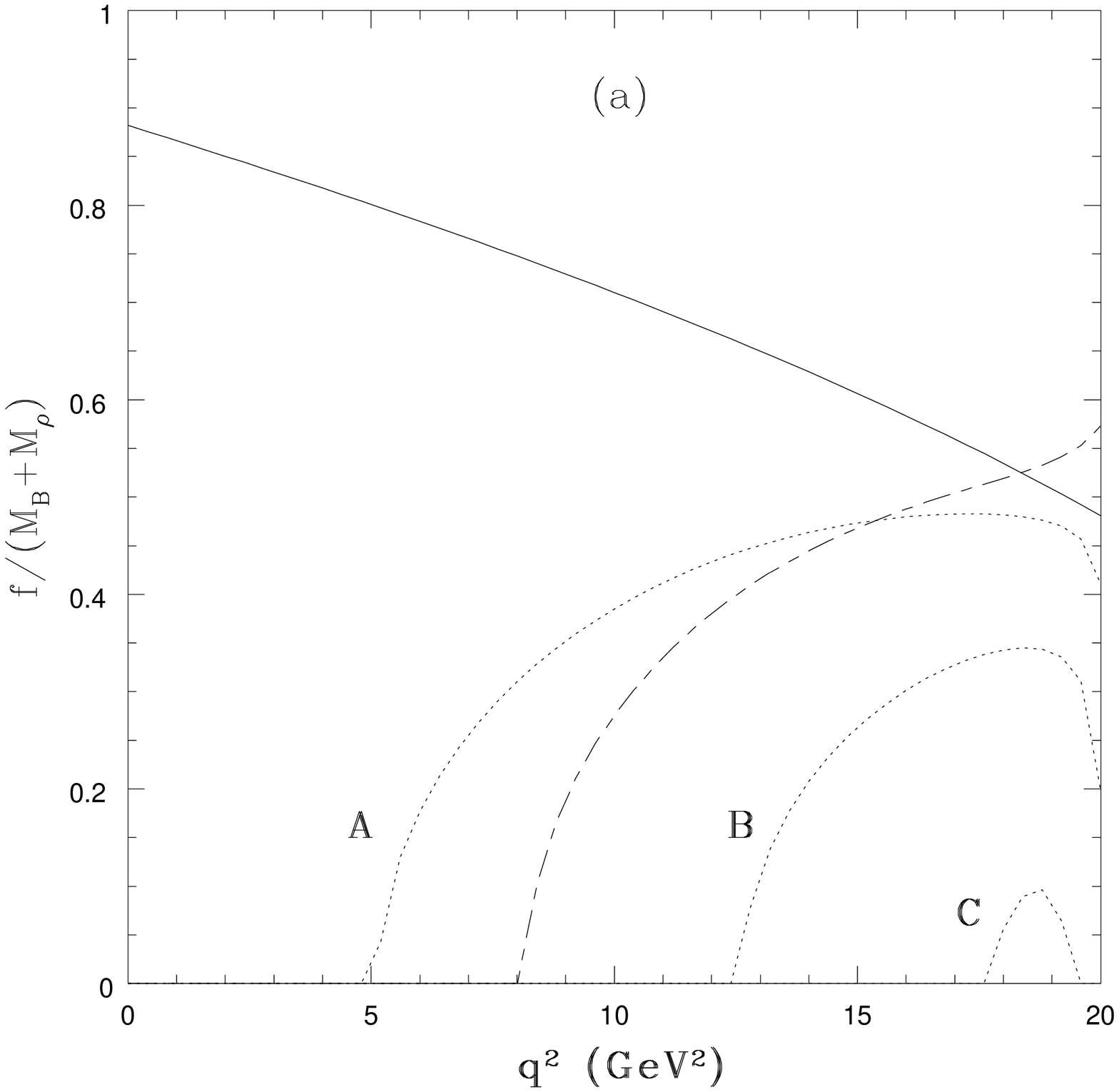}
\hfill
\epsfxsize=0.4\textwidth 
\epsffile{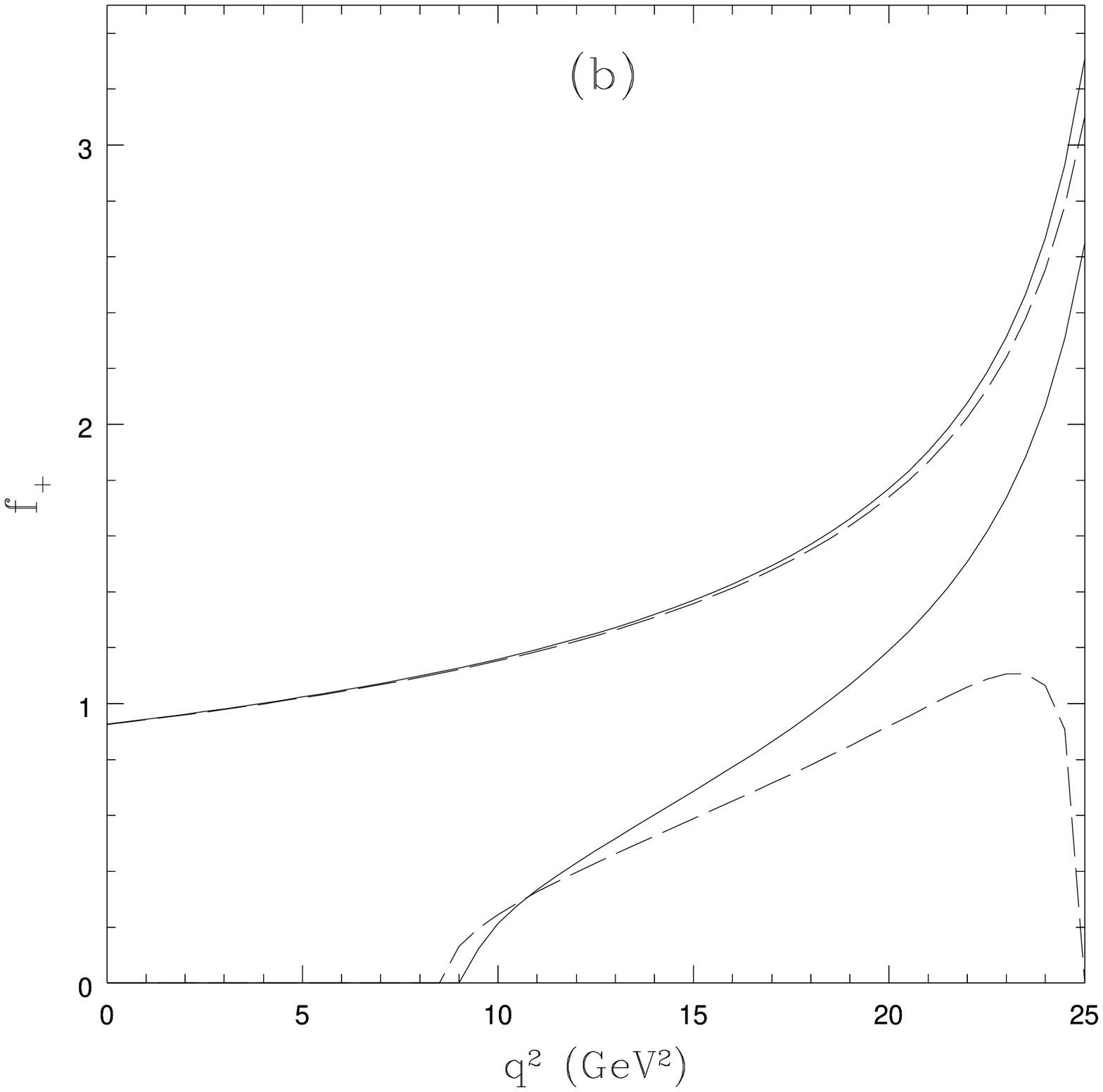}
\hfill} 
\caption{\it Upper and lower bounds on :  (a) The \btorhop\ form
factor $f(q^2)/(M_B + M_\rho)$.
Solid and dashed lines are bounds for $\Lam = 0.39\, {\rm GeV}, \lo= \lt=0$;
Dotted lines correspond to values given in the text.
(b) The  \btopip\ form factor $f_+(q^2)$.
Solid lines are for  $\Lam = 0.39, \lo=\lt=0$, dashed lines  
$\Lam = 0.39\, {\rm GeV}, 
\lo = -0.19 \,{\rm GeV}^2  , \lt = 0.12\, {\rm GeV}^2$.}
\label{fig1}
\end{figure}
%%%%%%%%%%%%%%%%%%%%%%%%%%%%%%%%%

Plotted in Fig.~1a are upper and lower bounds on the \btorhop\ form 
factor $f/(M_B + M_\rho)$ as a function of momentum transfer $q^2$. 
The lower bound, displayed for a range of
correlated $\Lam,\lo$ values taken from reference~\cite{GKLW2},
depends sensitively on the values of
$\Lam,\lo,$ and $\lt$, while the upper bound (solid line) has no dependence
on them at all. The dashed line is the lower bound  without
higher order corrections, using $\Lam = 0.39\, {\rm GeV}, \lo=0, \lt=0$.
We see that the bounds cross at
$q^2 \sim 18 \,{\rm GeV}$, indicating the need for higher order corrections.
The results of
including the $1/E$ corrections, using the measured value 
$\lt = 0.12 \,{\rm GeV}^2$, are illustrated by the dotted lines
$A,~B,~C$ choosing the values, $\left[ \Lam = 0.28 \,{\rm GeV}
,~\lo = -0.09 \,{\rm GeV}^2\right]$, the central values \cite{GKLW} 
$\left[ \Lam = 0.39 \,{\rm GeV}, ~\lo = -0.19\, {\rm GeV}^2\right]$ and
$\left[ \Lam = 0.50 \,{\rm GeV}, ~\lo = -0.29\, {\rm GeV}^2\right]$ 
respectively.
Clearly, the bounds are much more restrictive for low values of $\Lam$ and 
$|\lo|$. All the bounds except (C) are in a range relevant to ruling out typical 
models.
  
In Fig.~1b the upper and lower bounds on the \btopip\ form factor $f_+$ 
are plotted 
in solid lines 
for $\Lam = 0.39$ with vanishing $ \lo$ and $ \lt$ and in dashed lines
for $\Lam = 0.39\, {\rm GeV}, \lo = -0.19 \,{\rm GeV}^2  , \lt = 0.12\, {\rm 
GeV}^2$.
As with \btorhop\, the bounds are more restrictive for low values of
$\Lam$ and $|\lo|$. For example, the model of Wirbel, Stech, and 
Bauer\cite{WSB}\
is barely compatible with the lower dashed bound and is incompatible if
$\Lam=0.28\,{\rm GeV}^2$ is used instead (although no conclusions about
the reliability of such models can be made without the $\alpha_s$ corrections).

\begin{figure}
\centering
\epsfysize=2.5in   % Put x or y size here as epsfxsize of epsfysize
\hspace*{0in}
\epsffile{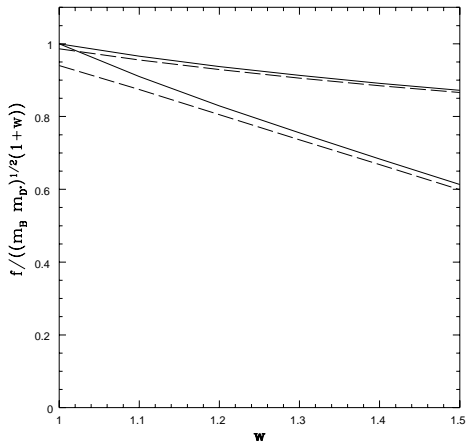}  % Put figure file title in here
\caption{\it Upper and lower bounds on the \btods\ form factor
$f/[(M_B M_{D^*})^{1/2}(1 + w)]$,  normalized to coincide with the Isgur-Wise
function in the heavy quark limit. Solid lines are for $\Lam = 0.39, \lo=0, 
\lt=0$,
dashed lines for $\Lam = 0.39\, {\rm GeV},\lo = -0.19\, {\rm GeV}^2,\lt = 0.12\, 
{\rm GeV}^2$.}

\label{fig3}  %put your label here which you can get with e.g. 
\end{figure}

Any bound with $m_q=0$ becomes unreliable when $E_H$ is too small, {\it i.e.}
when $q^2$ is too large. A hadronic energy greater than $\sim 1\, {\rm GeV}$, 
corresponding
to $q^2 \lesim 18\, {\rm GeV}$ for \btorho\ or \btopi\ , is probably necessary 
for
the $1/2 E_q$, $\alpha_s$, and local duality corrections to be under control.
When $m_q=m_c$, on the other hand, such corrections can be under control even
at zero recoil, $q_3 =0$. The bounds on the \btods\ form factor
$f/[ \sqrt{M_B M_{D^*} } (1 + w)]$, normalized to coincide with the Isgur-Wise
function in the infinite mass limit, are plotted against velocity transfer
$w = v \cdot v'$ in Fig.~3. Upper and lower bounds are shown in solid lines for
the leading order result $\Lam = 0.39, \lo=0, \lt=0$ and in dashed lines for 
$\Lam = 0.39\, {\rm GeV},\lo = -0.19\, {\rm GeV}^2,\lt = 0.12\, {\rm GeV}^2$.
Both sets of bounds are easily compatible with ALEPH\cite{ALEPH} data 
but only marginally compatible with CLEO\cite{CLEO} and DELPHI\cite{DELPHI} 
data.
Differentiating the upper and lower bounds at zero recoil with respect to $w$ 
leads 
to the generalizations of the Bjorken\cite{bj} and Voloshin\cite{Volo}
inequalities on the slope of the Isgur-Wise function. Perhaps even more  
interesting are the bounds on the normalization at zero recoil. 

The $\alpha_s$ and part of the $\alpha_s^2$ corrections to the upper bound
have been computed\cite{BSUV,GKLW,AC} but indicate the possibility of poor 
convergence for small $\Delta$. At order $\alpha_s$, the difference between the 
upper 
and lower bounds on $f(1)/[2 \sqrt{M_B M_{D^*}}]  $ is 
$ {-\alpha_s(\Delta) \over  \pi (M_1 - M_{D^*})} X_A^{(1)} $, where
\Eqa
X_A^{(1)} &=& {m_c (m_c+M_B)(m_c-3M_B)\over 9 M_B^2 }\ln{\Delta+m_c\over m_c}
\nonumber \\
&-&{\Delta \over 54 M_B^2 (\Delta+m_c)^2} \Biggl[6 m_c^4-12 m_c M_B \Delta^2
-18 m_c^2 M_B \Delta -12 m_c^3 M_B +9 m_c^3 \Delta
\nonumber \\
&-&14 \Delta^3 m_c-
10 m_c^2 \Delta^2-27 M_B^2 \Delta m_c-18 M_B^2 \Delta^2-4 \Delta^4-18 M_B^2 
m_c^2\Biggr].
\Endla{XAA}
Using $\Lam = 0.39\, {\rm GeV},\lo = -0.19\, 
{\rm GeV}^2,\lt = 0.12\, {\rm GeV}^2$ and a weight function 
$W_\Delta = \theta( \Delta-\eps)$ with $\Delta = 1 \,{\rm GeV}$ and 
$\alpha_s(\Delta) = 0.45$, we find that $\alpha_s$ 
corrections move the upper and lower bounds at threshold from
$ 0.99 \ge f(1)/[2\sqrt{M_B M_{D^*}}] \ge 0.94$ to
$ 0.96 \ge f(1)/[2\sqrt{M_B M_{D^*}}] \ge 0.90$. The upper bound
remains $0.96$ for higher $\Delta$ values,  $\Delta = 2 \,{\rm GeV}$ and 
$\alpha_s(\Delta) =
0.28$ or $\Delta = 3 \,{\rm GeV}$ and $\alpha_s(\Delta) = 0.23$, while the
lower bound decreases to $0.88$ or $0.86$ respectively. A better understanding
of the convergence properties in $\alpha_s$ is needed here.

\section{ Conclusions}

We have used inclusive sum rules to derive model-independent upper and 
lower bounds on  the form factors $f,g,a_+$ and $f_+$ 
for $\overline B \to D, D^*, \rho, \pi,  K$ and $ K^*$ semi-leptonic
and radiative decays, as well as upper and lower bounds
on the $B \to \omega$ form factor $f$. The method is 
easily generalized to other form factors or combinations of form factors 
and can be systematically improved by retaining higher order corrections
in $1/ 2 E_q$ or $\alpha_s$.  

At leading order, we find a surprising suppression of the \btorhop\
form factor $a_+$ at small momentum transfer, an experimentally 
verifiable prediction.
Other heavy-to-light form factors have upper and lower bounds that
are comparable to typical models.  We have included the leading 
$1/ 2 E_q$ nonperturbative corrections but not the  
$\alpha_s$ corrections.  This is generally sufficient for reliable upper bounds 
but not for lower bounds, which may be significantly modified by 
the $\alpha_s$ corrections. 

For \btods, we expect the $\alpha_s$ corrections to alter the bounds by only
a few percent, so these are reliable to that accuracy. We computed the 
$\alpha_s$ 
correction to the lower bound on the form factor $f(1)/2\sqrt{M_B M_{D^*}}$ at 
zero recoil. This widens the gap between the upper and lower bounds by only
$0.01$ for $\Delta = 1\, {\rm GeV}$. This should prove useful for 
extracting $V_{cb}$, as long as the $\alpha_s^2$ corrections
can be brought under control. 

The phenomenological implications of the sum rule bounds must await
a computation of the $\alpha_s$ corrections away from zero recoil.
We hope to present such an analysis in a later publication\cite{BR2}.
Once these terms are under control, the sum rule bounds may provide a means
to not only rule out various models, but also to constrain the values of 
the CKM elements $V_{cb}$, $V_{ub}$, and $V_{ts}$ from decays like
$\overline B \to D, D^*, l \overline \nu$, $\, 
\overline B \to \rho, \pi, l \overline \nu$, 
and $\overline B \to K^* \gamma$. For example, the \btorhop\ bounds
with $\Lam = 0.39\, {\rm GeV}, \lo=0,$ and $ \lt=0$, evaluated at $q^2= 12\, 
{\rm 
GeV}$, combined with a model-independent parameterization of $f$~\cite{BGL},
and lattice results at a single kinematic point $q^2\sim q^2_{max}$~\cite{latt}
constrain the total rate for \btorho\ to better than $40\%$.  Eventually,
it may be possible to forego lattice simulations in favor of experimental
data by using SU(3) and heavy quark symmetries\cite{LigWi}.

How good the constraints will be in reality depend crucially on the size and 
form of the $\alpha_s$ corrections as well as the actual values of
$\Lam$ and $\lo$.  The former can be addressed by explicit computation while 
the latter must await better experimental data (e.g., on the 
differential electron distribution in $\overline B \to X_s \gamma$). The 
possibility of making model-independent extractions of CKM elements 
like $V_{ub}$ and $V_{ts}$ is tantalizing and warrants continued investigation
in this area.

\vskip.2in {\centering\large\bf Acknowledgments}

We would like to thank Ben Grinstein and Mark Wise for useful discussions.
This work is supported in part by the Department of Energy under contract 
DOE-FG03-90ER40546. IZR would like to thank the Aspen Center for Physics
for its hospitality.

% ======================= BIBLIOGRAPHY ==========================

\newpage
\dspace{1.8}
\def\np#1{{\it Nucl. Phys.\ }{\bf #1}}
\def\pl#1{{\it Phys. Lett.\ }{\bf #1}}
\def\pr#1{{\it Phys. Rev.\ }{\bf #1}}


\begin{thebibliography}{99}                        

\bibitem{chay} J.~Chay, H.~Georgi and B.~Grinstein, Phys. Lett. {\bf
B247}, 399 (1990).

\bibitem{BSUV} B.~Bigi, M.~Shifman, N.~Uraltsev and A.I.~Vainshtein,
Phys. Rev. Lett. {\bf71}, 496 (1993).

\bibitem{BKSV} B.~Blok, L.~Koyrakh, M.~Shifman and A.I.~Vainshtein,
Phys. Rev. {\bf D49}, 3356 (1994).

\bibitem{MW} A.V.~Manohar and M.B.~Wise, Phys. Rev. {\bf D49}, 1310
(1994).

\bibitem{FLS} A.F.~Falk, M.E.~Luke and M.J.~Savage, Phys. Rev. {\bf
D49}, 3367 (1994).

\bibitem{Mann}T.~Mannel, Nucl. Phys. {\bf B413}, 396 (1994).

\bibitem{ar}
R. Akhoury and I.Z. Rothstein, hep-ph/9512303, to appear in {\it Phys.\ Rev.}
 {\bf D}.

\bibitem{BSUV2} B.~Bigi, M.~Shifman, N.~Uraltsev and A.I.~Vainshtein, 
Phys. Rev. {\bf D52}, 196 (1995).


\bibitem{GKLW}  B.~Grinstein, A.~Kapustin, Z.~Ligeti and M.~B.~Wise,
CALT-68-2041, hep-ph/9602262.
 
\bibitem{PQW} E.C.~Poggio, H.R.~Quinn and S.~Weinberg,
Phys. Rev. {\bf D13}, 1958 (1976).

\bibitem{hqet} N.~Isgur and M.~Wise, Phys. Lett. {\bf B232}, 113
(1989); E.~Eichten and B.~Hill, Phys. Lett. {\bf B234}, 511 (1990);
M. B.~Voloshin and M. A.~Shifman, Yad.\ Fiz.\ {\bf 47},
801 (1988) [Sov. J. Nucl. Phys. 47, 511 (1988)].

\bibitem{Volo} M.B. Voloshin, Phys. Rev. {\bf D46}, 3062 (1992).

\bibitem{BGM} C.G.~Boyd, B.~Grinstein and A.V.~Manohar, UCSD/PTH 95-19, to
appear in {\it Phys.\ Rev.} {\bf D}.

\bibitem{GKLW2}  B.~Grinstein, A.~Kapustin, Z.~Ligeti and M.~B.~Wise,
CALT-68-2043, hep-ph/9603314.

\bibitem{FLS2} A.F.~Falk, M.E.~Luke and M.J.~Savage, Phys. Rev. {\bf
D53}, 2491 (1996); Phys. Rev. {\bf D53}, 6316 (1996). 
 
\bibitem{akhster}
R. Akhoury and George Sterman, Phys.. Rev. {\bf D50}, 358 (1994).
\newline
R. Akhoury and I.Z. Rothstein, Phys. Lett. {\bf B337}, 176  (1994).
\newline
M. Dahm, R. Jakoband  P. Kroll, {\bf Z} .Phys. {\bf C68}, 595, (1995).

\bibitem{WSB} M. Bauer, B. Stech  and  M. Wirbel,  Z. Phys. {\bf C34},
 103 (1987).

\bibitem{ALEPH} D.~Buskulic \etal\ (ALEPH Collaboration), Phys.
Lett. {\bf B359} 236 (1995).

\bibitem{CLEO} B.~Barish \etal\ (CLEO  Collaboration), Phys. Rev. {\bf D51},
1014 (1995).

\bibitem{DELPHI} P.~Abreau \etal (DELPHI  Collaboration), CERN-PPE/96-11.

\bibitem{bj} J.D.~Bjorken, invited talk at Les Rencontres de Physique
de la Vallee d'Aoste, La Thuile, Italy, SLAC Report No. SLAC-PUB-5278
(1990), unpublished;\\ N.~Isgur and M.~Wise, Phys. Rev. {\bf D43}, 819
(1991);\\ J.D.~Bjorken, I.~Dunietz and J.~Taron, Nucl. Phys. {\bf
B371}, 111 (1992).

\bibitem{AC} A.~Czarnecki, TTP96-05, hep-ph/9603261.

\bibitem{BR2} C.G.~Boyd, I.~Rothstein, in preparation.

\bibitem{BGL} C.G.~Boyd, B.~Grinstein and R.F.~Lebed, Phys. Rev. Lett.
{\bf74}, 4603 (1995);  Phys. Lett. {\bf B353}, 306 (1995);
Nucl. Phys. {\bf B461}, 493 (1996); C.G.~Boyd and R.F.~Lebed, UCSD-PTH-95-23,
hep-ph/9512363.

\bibitem{latt} J.M.~Flynn \etal (UKQCD Coll.), Nucl. Phys. {\bf B461},
327 (1996).

\bibitem{LigWi}  Z.~Ligeti and M.~B.~Wise, CALT-68-2029, hep-ph/9512225.

\end{thebibliography}
\end{document}